\documentclass{ws-p10x7}
\newcommand {\qq} {\mathrm{q_{tag}}\overline{\mathrm{q}}_{\mathrm{tag}}}
\newcommand {\gincl} {\mathrm{g}_{\mathrm{incl.}}}
\newcommand {\epem}   {e$^+$e$^-$}

\newcommand {\xpmin} {x_p^{min.}}
\newcommand {\cacf} { \mathrm{C}_{\mathrm{A}}/\mathrm{C}_{\mathrm{F}} }

\begin{document}

\title{ \hspace*{9.8cm} {\normalsize UCRHEP-E288} \\
\hspace*{10.3cm}{\normalsize August 21, 2000} \\[1cm]
Jet Structure Studies at LEP and HERA$^\dagger$}

\author{J. William Gary }

\address{Department of Physics, University of California, 
Riverside CA 92521, USA\\
E-mail: bill.gary@ucr.edu}

\twocolumn[\maketitle\abstract{
A summary
of some recent studies in jet physics is given.
Topics include leading particle production in light flavor
events in e$^+$e$^-$ annihilations,
an analytical treatment of gluon and quark jets at the
next-to-next-to-next-to-leading order (3NLO),
and various studies performed at LEP and HERA involving
separated gluon and quark jets.
}]

\footnotetext{$^\dagger$Talk presented at the XXXth International 
Conference on High Energy Physics, July 27 - August 2, 2000, 
Osaka, Japan.}

\section{Leading particle production in separated light
quark events}
\label{sec-leading}

Separated charm (c) and bottom (b) quark events have
been well studied in e$^+$e$^-$ annihilations.
c and b quarks are mostly produced
at the electro-weak vertex in
e$^+$e$^-\rightarrow\mathrm{q}\overline{\mathrm{q}}\rightarrow hadrons$
events,
in which the primary quark $\mathrm{q}$ is c or b,
and are relatively easy to identify.
In contrast, separated up (u), down (d) and strange (s)
events have not been much studied.
u, d and s quarks are copiously produced during jet
development,
making events in which they are produced as primary
quarks more difficult to identify.

A recent study\cite{bib-opalleading} by the OPAL Collaboration
at LEP identifies
e$^+$e$^-\rightarrow\mathrm{q}\overline{\mathrm{q}}\rightarrow hadrons$
events with $\mathrm{q}$=u, d or s.
The probabilities $\eta_{\mathrm{q}}^i(x_p^{min.})$
are determined for the quark q=u, d, s (or charge conjugate)
to appear in an identified hadron $h_i(x_p^{min.})$,
where $h_i$=$\pi^+$, $K^+$, $K^0_S$, $p$=proton, or $\Lambda$
(or charge conjugate) is the leading particle in the jet,
i.e.~it has the largest momentum,
with a scaled momentum $x_p$=$2p/\sqrt{s}$ larger
than a minimum $x_p>x_p^{min.}$,
where $\sqrt{s}$=91~GeV.
The $\eta_{\mathrm{q}}^i$ probabilities provide
unique, detailed information on the hadronization process
and a more direct determination of basic hadronization parameters
such as the strange quark suppression factor
$\gamma_s\equiv\,$Prob.(s)/Prob.(u or d)
than in most previous studies.

The method for the measurement is presented
in ref.\cite{bib-letts}.
Briefly, e$^+$e$^-\rightarrow hadrons$ events are divided
into hemispheres using the plane perpendicular to the
thrust axis.
Thus an {\it inclusive} definition of jets is used,
i.e. a jet is an event hemisphere.
The single and double tag rates in the jets are measured,
$N_i(\xpmin)/N_{had.}$ and $N_{ij}(\xpmin)/N_{had.}$,
where $N_i(\xpmin)$ is the number of jets in which the highest
momentum particle has $x_p>\xpmin$ and is identified as
$\pi^+$, $K^+$, $K_S^0$, $p$ or $\Lambda$,
and $N_{ij}(\xpmin)$ is the analogous quantity for events in
which a hadron $h_i$ is identified in one hemisphere and 
hadron a $h_j$ is identified in the other.

The system of equations
\begin{center}
 \begin{tabular}{lcl}
   {\large $\frac{ N_i(\xpmin) }{ N_{had.} }$} 
    & = & 2
      ${\displaystyle \sum_{q=u,d,s,c,b} }
      \eta_{\mathrm{q}}^i(\xpmin) \mathrm{R}_{\mathrm{q}}$ \\
   {\large $\frac{ N_{ij}(\xpmin) }{ N_{had.} }$ }
    &= & (2$\,-\delta_{ij}) \times $ \\
 \multicolumn{3}{c}{ $ {\displaystyle \sum_{q=u,d,s,c,b}}
     \rho_{ij}(\xpmin)  \eta_{\mathrm{q}}^i(\xpmin)
        \eta_{\mathrm{q}}^i(\xpmin) \mathrm{R}_{\mathrm{q}}$
 }
\end{tabular}
\end{center}
is then solved,
where $\mathrm{R}_{\mathrm{q}}$=$ {\displaystyle\frac{
           \Gamma_{\mathrm Z^0\rightarrow q\overline{q}}}
          {\Gamma_{Z^0\rightarrow hadrons}} }$
is the partial decay width of the Z$^0$ to the different
quark flavors, taken from the standard model for 
q=u, d, s and from LEP for q=c, b,
and $\rho_{ij}$ are hemisphere correlations due mostly
to well understood effects such as geometric acceptance
and gluon bremsstrahlung.
The $\rho_{ij}$ factors are evaluated using
QCD Monte Carlo and are the only MC information entering
the equations.  
Their values are typically $\rho_{ij}$$\approx$1.01-1.10.
To solve the above system of equations,
supplementary information involving assumptions of
isospin symmetry and the flavor independence of the
strong interaction are invoked 
(see refs.\cite{bib-opalleading,bib-letts}).

\begin{figure}
  \epsfxsize200pt
  \figurebox{}{}{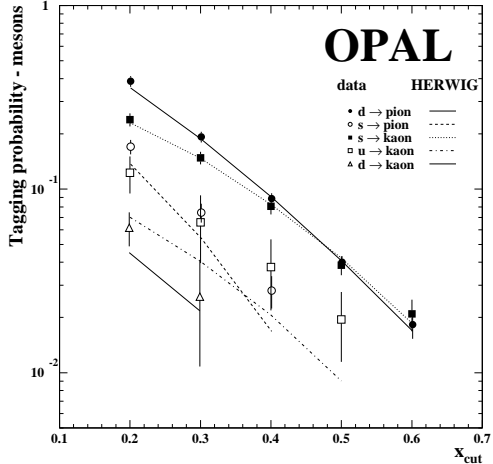}
\caption{Tagging probabilities for pions and kaons
as a function of the minimum scaled momentum~$\xpmin$,
in comparison to Monte Carlo predictions.}
\label{fig-leading}
\end{figure}

The measured probabilities $\eta_{\mathrm{q}}^i$ for 
$i$=$\pi$ and $K$ are shown in figure~\ref{fig-leading} 
as a function of $\xpmin$.
From this figure,
the so-called leading particle effect is clearly visible,
i.e.~primary quarks appear primarily as valence quarks
in the highest momentum hadrons.
Thus d and u quarks appear predominantly in pions,
and s quarks in kaons,
rather than vice versa.
The leading particle effect has been observed many times
for b and c quarks.
There is only one previous study of the effect for
light quarks,
given in ref.\cite{bib-sldleading}.

\begin{figure}
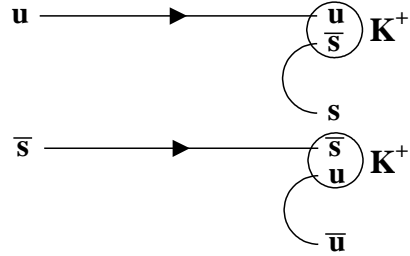

   \begin{center}
    \begin{tabular}{c}
  \epsfxsize150pt
      \figurebox{}{}{gammas_u_to_kplus_bw.epsi} \\
  \epsfxsize150pt
      \figurebox{}{}{gammas_s_to_kplus_bw.epsi} \\
    \end{tabular}
   \end{center}
\caption{Illustration of the determination of the
strangeness suppression factor $\gamma_s$ from the
ratio of the tagging probabilities for $K^{\pm}$ mesons
to be produced either from a primary u quark (top)
or primary $\overline{\mathrm{s}}$ quark (bottom).}
\label{fig-gammas}
\end{figure}

\mbox{Using the probabilities $\eta_{\mathrm{q}}^i$,
basic hadron-} ization
parameters such as the strange
quark suppression factor $\gamma_s$ can be determined
rather directly.
For example the probability factors
$\eta_{\mathrm{u}}^{K^{\pm}}$ and
$\eta_{\mathrm{s}}^{K^{\pm}}$
differ only by $\mathrm{s}\overline{\mathrm{s}}$
or $\mathrm{u}\overline{\mathrm{u}}$ pair production
from the sea,
see figure~\ref{fig-gammas}.
Their ratio thus determines $\gamma_s$.
Note that most other measurements of $\gamma_s$
either compare the yields of hadrons with different masses,
such as kaons and pions,
or else rely on the tuning of Monte Carlo parameters.
They are thus not as direct as the method described here
which employs kaons only.
The result for $\gamma_s$ is 
$0.422\pm0.049\,(\mathrm{stat.})\pm 0.059\,(\mathrm{syst.})$,
about one standard deviation higher than the
result in ref.\cite{bib-sldleading}
based on a similar technique.

\section{Experimental properties of gluon and quark jets
from a point source}

The inclusive (hemisphere) definition of jets yields
unbiased jets whose properties can be compared directly 
to the predictions of analytic calculations.
Inclusive production of hadrons in {\epem} annihilations
provides a natural source for unbiased quark jets,
used
--~for example~--
in the study described in section~\ref{sec-leading}.
One can ask if an analogous sample of high energy unbiased
gluon jets can be identified.

The answer,
as discussed in ref.\cite{bib-jwg94}, is yes,
by selecting rare events in {\epem} annihilations
in which two identified quark jets appear in the
same hemisphere of an event.
The opposite hemisphere,
against which the two quark jets recoil,
approximates an unbiased gluon
jet with high accuracy\cite{bib-jwg94}.
Such events have been labeled
e$^+$e$^-\rightarrow\qq\gincl$ events
to differentiate them from ``ordinary''
$\mathrm{q}\overline{\mathrm{q}}$g three-jet events
defined using a jet finder.
The tagged quark jets 
$\mathrm{q_{tag}}$ and $\overline{\mathrm{q}}_{\mathrm{tag}}$
are identified using b tagging.
The recoiling hemisphere ``$\gincl$'' is the
unbiased gluon jet.

Experimental analysis of $\gincl$
jets from Z$^0$ decays has been
presented by OPAL\cite{bib-opalglincl}.
Starting from their full LEP-1 data sample of about 4 million events,
a sample of 439 $\gincl$ jets
is isolated with a purity of about 82\%.
The $\gincl$ jet energy is about 40~GeV.
The $\gincl$ jets are compared
to a sample of light (u,d and s) quark jets,
also from Z$^0$ decays,
defined using the hemisphere definition.
The quark jet sample is resticted to light flavors
to better approximate the massless quark condition employed
by analytic calculations.

Only one aspect of the results will be presented here,
namely the ratio of soft particles at large transverse momentum
$p_\perp$ between the unbiased gluon and quark jets.
$p_\perp$ is defined with respect to the jet axis.
As discussed in ref.\cite{bib-khoze98},
this ratio provides a direct measurement
of the QCD color factor ratio $\cacf$.

\begin{figure}
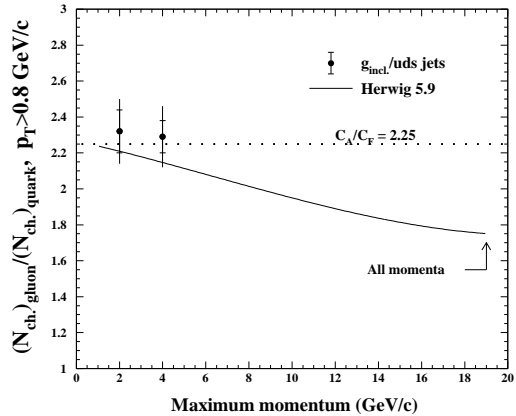

  \epsfxsize190pt
  \figurebox{}{}{fig_pt_gt0p8_bw.epsi}
\caption{Ratio $r$ of charged particle multiplicities between
unbiased gluon and quark jets for particles with large transverse
momenta to the jet axis defined by $p_\perp$$\,>\,$0.8~GeV/$c$.
The results are shown as a function of the softness of the
the particles,
defined by the maximum particle momentum used
to determine~$r$.
}
\label{fig-pt}
\end{figure}

Figure~\ref{fig-pt} shows the charged particle multiplicity ratio
between the unbiased gluon and quark jets, $r$,
as a function of the softness of the particles.
The softness of the particles is
defined by the maximum particle momentum $p_{\mathrm{max.}}$
considered when determining~$r$.
The particles are required to have $p_\perp$$\,>\,$0.8~GeV/$c$.
Particles with $p_\perp$$\,<\,$0.8~GeV/$c$
are dominated by the effects of hadronization
as is established using MC.
The solid curve shows the prediction of the Herwig Monte Carlo.

With no explicit cut on $p_{\mathrm{max.}}$
(``All momenta'')
the multiplicity ratio is predicted to be about~1.8.
As softer and softer particles are selected
($p_{\mathrm{max.}}$ is decreased),
the curve approaches the QCD result
$\cacf$=2.25 as predicted in ref.\cite{bib-khoze98}.
OPAL results are shown for $p_{\mathrm{max.}}$=2~GeV/$c$
and 4~GeV/$c$.
The result using $p_{\mathrm{max.}}$=4~GeV/$c$
is $r$=$2.29\pm0.017 \;\mathrm{(stat.+syst.)}$
which provides one of the most accurate current
experimental determinations of $\cacf$.
Note that unlike \underline{\it all other methods},
this result is not based on a fit of a QCD motivated 
expression
--~in which $\cacf$ is extracted as a fitted
parameter~--
but is the ratio of directly measured quantities.

\section{\mbox{Analytic description of multipli-}
city in gluon and quark jets}
\label{sec-analytic}

An analytic description of multiplicity in unbiased gluon 
and quark jets has recently been performed at
the next-to-next-to-next-to-leading order (3NLO)
in perturbation theory\cite{bib-dremingary}.
Details of the calculation are presented 
in refs.\cite{bib-dremingary,bib-dremintran}.
Here some comparisons of the results with experiment
will be discussed.

\begin{figure}
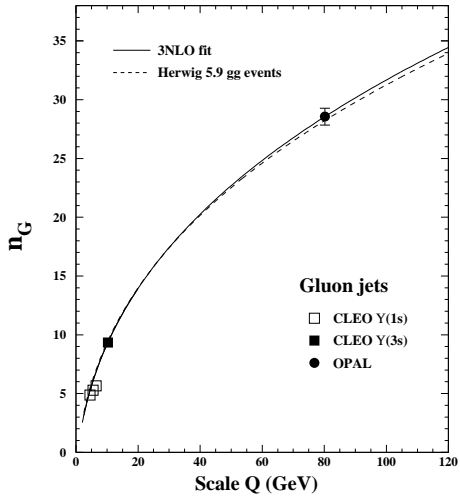

  \epsfxsize170pt
  \figurebox{}{}{nch_gluon_vs_scale_bw.epsi}
\caption{Measurements of unbiased gluon jet  charged
particle multiplicity
in comparison to Monte Carlo and the 3NLO analytic predictions.}
\label{fig-gluons}
\end{figure}

The currently available measurements of
mean multiplicity in unbiased gluon jets, $n_G$,
from the CLEO Collaboration at CESR\cite{bib-cleo}
at low energies and from OPAL $\gincl$ jets\cite{bib-opalglincl}
at high energy,
are shown in figure~\ref{fig-gluons}.
The dashed curve is the prediction of Herwig,
which describes the data rather well.
The solid curve is a fit of the 3NLO analytic 
prediction\cite{bib-dremingary}
using two free parameters:
(1)~an overall normalization K and (2)~an effective QCD
scale parameter~$\Lambda$.
Translating the fitted result for $\Lambda$ into
$\alpha_S$(M$_{\mathrm{Z}}$) yields
$\alpha_S$(M$_{\mathrm{Z}}$)=$0.14\pm0.01$,
not too different from the world average value
$\alpha_S$(M$_{\mathrm{Z}}$)$\approx$0.12
found using $\Lambda_{\overline{\mathrm{MS}}}$.

\begin{figure}
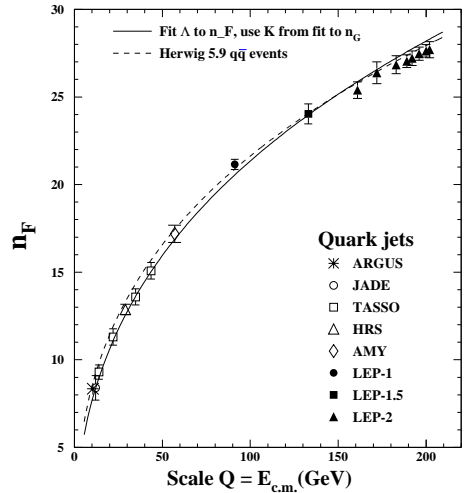

  \epsfxsize170pt
  \figurebox{}{}{nch_quark_vs_scale_bw.epsi}
\caption{Measurements of unbiased quark jet charged
particle multiplicity
in comparison to Monte Carlo and the 3NLO analytic predictions.}
\label{fig-quarks}
\end{figure}

Measurements of mean multiplicity in unbiased quark jets,
$n_F$,
performed by many experiments,
are shown in figure~\ref{fig-quarks}.
Again Herwig (dashed curve) describes the
data rather well.
Making a one parameter fit of the 3NLO 
expression for quark jet multiplicity\cite{bib-dremingary}
to the data (solid curve),
with $\Lambda$ as the fitted parameter
and with the normalization K fixed
from the fit to the gluon jet data,
yields $\alpha_S$(M$_{\mathrm{Z}}$)=$0.135\pm0.002$,
not very different from the result presented above 
for gluon jets.
This demonstrates the consistency of the analytic
approach to the growth of multiplicity with scale.
Note that the inclusive (unbiased) definition of jets
is crucial to obtain this result.
If the quark and gluon jets are defined using a 
jet finder such as the $k_\perp$ jet finder,
such consistency is not observed
(see ref.\cite{bib-dremingary}).

\begin{figure}
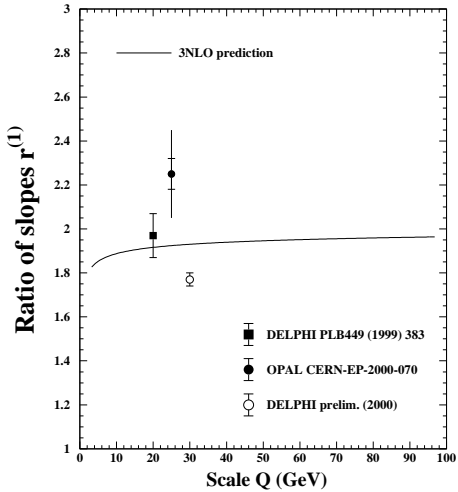

  \epsfxsize170pt
  \figurebox{}{}{slopes_r1_bw.epsi}
\caption{The ratio of slopes $r^{(1)}$ between gluon
and quark jets in comparison to the 3NLO analytic
prediction.}
\label{fig-slopes}
\end{figure}

One other result will be discussed here,
namely the ratio of the {\it slopes} of multiplicity,
defined by 
$\displaystyle r^{(1)} = \frac{ d\langle n_G\rangle / dy }
{ d\langle n_F\rangle / dy }$
where $y$=$\ln (Q/\Lambda)$ with $Q$ the jet energy.
The ratio of slopes $r^{(1)}$ has the same asymptotic
limit of 2.25 as $r$,
but is predicted\cite{bib-dremingary}
to have smaller pre-asymptotic corrections.
The 3NLO prediction for $r^{(1)}$ versus $Q$ is shown
in figure~\ref{fig-slopes}
in comparison to measurements from OPAL\cite{bib-opaleta}
and the DELPHI\cite{bib-delphir1} Collaboration at LEP.
The experimental results exhibit some scatter because
of differences in the definition of jets but are in
general agreement with the prediction $r^{(1)}$$\approx$1.9.

\section{Substructure dependence of dijet cross sections
in photoproduction at HERA}

A recent study
from the ZEUS Collaboration at HERA
concerns the photoproduction of dijets 
in low Q$^2$ ep scattering\cite{bib-zeusjets}.
Specifically,
the reaction
$\gamma$p$\,\rightarrow\,2\,jets\,+\,X$ is studied,
where the two-jet system is either
$\mathrm{q}\overline{\mathrm{q}}$, gg or $\mathrm{q}$g.
Jet shapes (see below) are used to tag samples
enhanced in quark and gluon jets.
Measurements are then made of the rapidity distributions
of the jets.

Events with at least two jets are selected using the
longitudinally invariant $k_\perp$ jet finder.
Events are retained if they contain at least two jets
with transverse energy E$_{\mathrm{T}}$$>$14~GeV in the
pseudo-rapidity range \mbox{--1}$<$$\eta$$<$2.5,
where $\eta$=--$\ln(\tan(\theta/2))$, with 
$\theta$ the polar angle with respect to the 
proton beam direction.
The two jets with highest E$_{\mathrm{T}}$ are analyzed.
Specifically, the jet profiles and sub-jet multiplicities
of these jets are determined.
The jet profile $\Psi(r)$ is the distribution of the
fraction of jet energy inside a cone of half radius
$r$ around the jet axis,
relative to a cone with $r$=1~radians.

\begin{figure}
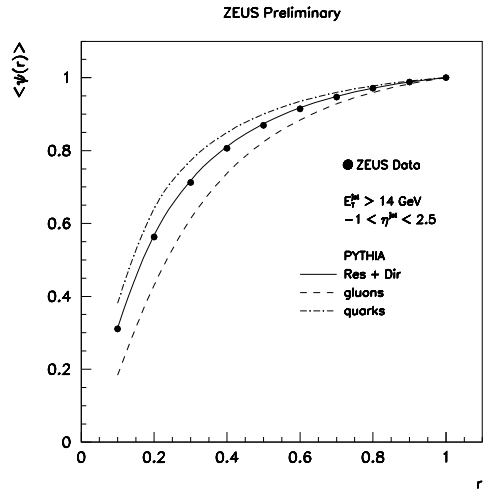

  \epsfxsize180pt
  \figurebox{}{}{zeus_profiles_bw.epsi}
\caption{Jet profile of the two highest energy jets in
$\gamma$p$\,\rightarrow\,2\,jets\,+\,X$ events.}
\label{fig-profile}
\end{figure}

The mean profile of the selected jets is
presented in figure~\ref{fig-profile}.
The data, shown by the points,
are well represented by the Monte Carlo,
indicated by the solid line.
The individual contributions of gluon and quark jets
as predicted by the simulation 
are shown by the dashed and dash-dotted curves,
respectively.
The profiles of the gluon and quark jets
are seen to differ substantially.
In particular,
gluon jets are predicted to be much less collimated 
around the jet axis than quark jets,
a fact that has been well established experimentally
in {\epem} collisions\cite{bib-opal93}.

\begin{figure}
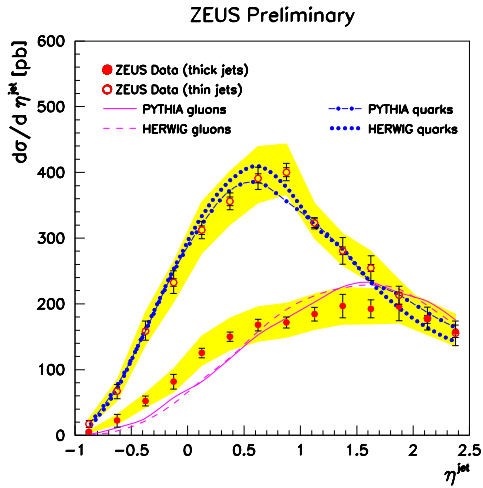

  \epsfxsize180pt
  \figurebox{}{}{zeus_eta_data.epsi}
\caption{Rapidity distribution of ``thick'' and ``thin'' jets
in comparison to the Monte Carlo predictions for
gluon and quark jets in $\gamma$p collisions.
}
\label{fig-rapidity}
\end{figure}

Choosing a cone size $r$=0.3 radians,
quark and gluon jet dominated samples
are selected by requiring 
$\Psi(r$=0.3$)$$>$0.8 or $\Psi(r$=0.3$)$$<$0.6,
respectively.
The resulting samples are denoted ``thin'' jets and
``thick'' jets and have quark and gluon jet purities of
about 85\% and~60\%, again respectively.
The rapidity distributions of ``thin''
and ``thick'' jets are shown in 
figure~\ref{fig-rapidity}
by the open and solid points.
The curves show the Monte Carlo predictions
for quark and gluon jets in $\gamma$p collisions.
Gluon jets are produced mostly through {\it resolved}
processes,
in which the photon in the $\gamma$p collision
acts as a source of partons.
In contrast,
quark jets are produced mostly through {\it direct} 
processes in which the photon couples directly 
to partons in the proton.
This explains the different rapidity distributions
predicted for gluon and quark jets.
The ``thick'' and ``thin'' jet measurements are seen
to follow the predictions for gluon
and quark jets quite well,
demonstrating a successful separation and test
of the cross sections for gluon and quark jets
individually.

\section{$\pi^0$, $\eta$, $K^0$ and charged particle
multiplicities in quark and gluon jets}

Last,
I discuss a recent study from OPAL\cite{bib-opaleta}
on differences in the production rates of identified 
particles in gluon and quark jets.

QCD predicts that $r_h$ 
--~the ratio of the mean multiplicities
of identified hadrons between gluon and quark jets~--
is the same for all hadrons~$h$.
In practice,
$r_h$ might differ for different particles because of
the decay properties of hadrons or because of dynamical
differences between the hadronization of gluons and quarks.
For example,
the gluon octet model of Peterson and Walsh\cite{bib-peterson}
predicts $r_h$ to be larger for isoscalar particles than
for non-isoscalars,
e.g. an enhancement in $r_h$ for $\eta$ mesons
compared to charged particles.
So far,
studies at LEP of identified particles in gluon jets
have either considered
(1)~$\pi^+$, $K^+$, $p$, $K_S^0$ and $\Lambda$
production (or c.c.)
in identified samples of gluon and quark 
jets\cite{bib-leprh},
leading to experimental determinations of $r_h$,
or else
(2)~the production of
$\pi^0$, $\eta$, $\eta^\prime$, $K_S^0$ and $\Lambda$
(or c.c.)
in the lowest energy jet of {\epem} three-jet 
events in comparison to the corresponding
result for charged particles or to Monte Carlo 
predictions\cite{bib-l3k0s,bib-l3eta,bib-alepheta}.
The OPAL study discussed here is the first to employ
the former strategy for $\pi^0$ and $\eta$ mesons,
i.e. the first to determine
$r_{h}$ for $\pi^0$ and~$\eta$.

Three-jet events are selected using the $k_\perp$,
cone, or LUCLUS jet finders:
the difference in the results defines a systematic uncertainty
related to the jet definition.
The jets are ordered such that jet~1 has
the highest energy and jet~3 the lowest.
The jet energy specifies the gluon jet content of the jet.
The jets are then examined in terms of the so-called
hardness scale\cite{bib-hardness}
Q$_{\mathrm{jet}}$=E$_{\mathrm{jet}}
\sin(\theta_{\mathrm{min.}}/2)$,
where $\theta_{\mathrm{min.}}$ is the smaller of the
angles with respect to the two other jets.
The hardness scale has been shown\cite{bib-delphir1}
to be a much more appropriate scale than the jet energy
E$_{\mathrm{jet}}$ when
comparing jets embedded in a three-jet environment,
i.e.~when comparing biased jets,
as opposed to the unbiased jets discussed in
sections~\ref{sec-leading}-\ref{sec-analytic}.

The $\pi^0$, $\eta$, $K^0_S$ and charged particle rates
in jets 2 and 3 are compared in their overlap region,
defined by 7$<$Q$_{\mathrm{jet}}$$<$30~GeV.
The multiplicity measurements are unfolded
algebraically using the known quark and gluon jet
content of jets 2 and~3 to obtain results corresponding 
to pure gluon and quark jets.
The results for $r_h$ for $\pi^0$, $\eta$
and $K_S^0$ are then divided by the corresponding
result for charged particles to obtain:
\begin{center}
 \begin{tabular}{ccc}
      $r_{\eta}/r_{ch.}$  & = & $1.09\pm 0.12$ \\
      $r_{\pi^0}/r_{ch.}$ & = & $1.01\pm 0.04$ \\
      $r_{K^0_s}/r_{ch.}$ & = & $0.95\pm 0.04$
 \end{tabular}
\end{center}
All three results are consistent with unity,
indicating no evidence for a dynamical difference
in the hadronization of gluon and quark jets.
This conclusion is in agreement with recent results
from the ALEPH Collaboration at LEP\cite{bib-alepheta}.
The OPAL and ALEPH results for $\eta$ mesons
contradict the conclusion of an
earlier study\cite{bib-l3eta}
by the L3 Collaboration at LEP,
in which evidence for a dynamical enhancement of
$\eta$ mesons in gluon jets was reported.

\end{document}